\pdfoutput=1

\documentclass[11pt,twoside,a4paper,cmspaper,final,collab]{cms-tdr}
\def\svnVersion{440451P}\def\cmsCernNoTag{CERN-EP-2017-069}\def\cmsCernDate{\today}\def\cmsMessage{Published in Physics Letters B as \href{http://dx.doi.org/10.1016/j.physletb.2017.11.043}{\doi{10.1016/j.physletb.2017.11.043.}}}

\begin{document}\cmsNoteHeader{TOP-16-010}

\hyphenation{had-ron-i-za-tion}
\hyphenation{cal-or-i-me-ter}
\hyphenation{de-vices}
\RCS$Revision: 435100 $
\RCS$HeadURL: svn+ssh://svn.cern.ch/reps/tdr2/papers/TOP-16-010/trunk/TOP-16-010.tex $
\RCS$Id: TOP-16-010.tex 435100 2017-11-18 21:59:11Z alverson $
\newlength\cmsFigWidth
\ifthenelse{\boolean{cms@external}}{\setlength\cmsFigWidth{0.85\columnwidth}}{\setlength\cmsFigWidth{0.4\textwidth}}
\ifthenelse{\boolean{cms@external}}{\providecommand{\cmsLeft}{top\xspace}}{\providecommand{\cmsLeft}{left\xspace}}
\ifthenelse{\boolean{cms@external}}{\providecommand{\cmsRight}{bottom\xspace}}{\providecommand{\cmsRight}{right\xspace}}

\newcommand{\mue}{\ensuremath{\PGm^\pm\Pe^\mp}\xspace}
\newcommand{\emu}{\ensuremath{\Pe^\pm\PGm^\mp}\xspace}
\newcommand{\ee}{\ensuremath{\Pe^\pm\Pe^\mp}\xspace}
\newcommand{\mumu}{\ensuremath{\PGm^\pm \PGm^\mp}\xspace}
\newcommand{\ttbb}{\ensuremath{\ttbar\bbbar}\xspace}
\newcommand{\ttcc}{\ensuremath{\ttbar\PQc\PAQc}\xspace}
\newcommand{\ttb}{\ensuremath{\ttbar\PQb\cmsSymbolFace{j}}\xspace}
\newcommand{\ttll}{\ensuremath{\ttbar\cmsSymbolFace{LF}}\xspace}
\newcommand{\ttjj}{\ensuremath{\ttbar\cmsSymbolFace{jj}}\xspace}
\newcommand{\ttH}{\ensuremath{\ttbar\PH}\xspace}

\newcommand{\sttbb}{\ensuremath{\sigma_{\ttbb}}\xspace}
\newcommand{\sttjj}{\ensuremath{\sigma_{\ttjj}}\xspace}

\cmsNoteHeader{TOP-16-010}
\title{Measurements of $\ttbar$ cross sections in association with b jets and inclusive jets and their ratio using dilepton final states in pp collisions at $\sqrt{s} = 13$\TeV}

\date{\today}

\abstract{
The cross sections for the production of $\ttbb$ and $\ttjj$ events and their ratio $\sigma_{\ttbar\bbbar}/\sigma_{\ttjj} $ are measured using data corresponding to an integrated luminosity of 2.3\fbinv collected in pp collisions at $\sqrt{s} = 13$\TeV with the CMS detector at the LHC.
Events with two leptons ($\Pe$ or $\PGm$) and at least four reconstructed jets, including at least two identified as b quark jets, in the final state are selected.
In the full phase space, the measured ratio is $0.022 \pm 0.003\stat\pm0.006\syst$,
the cross section $ \sigma_{\ttbar\bbbar}$ is $4.0 \pm 0.6\stat\pm1.3\syst$\unit{pb} and $ \sigma_{\ttjj}$ is $184 \pm 6\stat\pm 33\syst$ \unit{pb}.
The measurements are compared with the standard model expectations obtained from a \POWHEG simulation at next-to-leading-order interfaced with \PYTHIA.
}

\hypersetup{%
pdfauthor={CMS Collaboration},%
pdftitle={Measurements of ttbar cross sections in association with b jets and inclusive jets and their ratio using dilepton final states in pp collisions at sqrt(s) = 13 TeV},%
pdfsubject={CMS},%
pdfkeywords={CMS, physics, top quark}}

\maketitle

\section{Introduction}\label{sec:intro}

Since the discovery of the Higgs boson~\cite{atlas:newhiggs,cms:newhiggs,cms:newhiggsdetail}, its properties have been measured and compared to the standard model (SM) prediction~\cite{cms:propertieshiggs_fourlepton,cms:higgstoWW,cms:Htautau,cms:VHbb,atlas:newhiggscoupling,atlas:newhiggsevidence}.
However, the coupling of the top quark to the Higgs boson remains to be determined.
Although it appears indirectly through loops in the gluon--gluon fusion production process and in the $\PH\to\gamma\gamma$ decay channel, a direct measurement has yet to be completed.
One of the most promising channels for a direct measurement of the top quark Yukawa coupling in the SM is the production of the Higgs boson in association with a \ttbar pair (\ttH), where the Higgs boson decays to \bbbar, thus  leading to a \ttbb final state.
This final state, which has not been observed yet~\cite{combination:higgs}, has an irreducible nonresonant background from the production of a top quark pair in association with a b quark pair produced via gluon splitting ($\Pg \to\bbbar$).

Calculations of the inclusive production cross section for \ttbar events with additional jets have been performed to next-to-leading-order (NLO) precision for proton--proton centre-of-mass energies of 7, 8, and 13\TeV~\cite{ttbbQCD}.
The dominant uncertainties in these calculations are from the choice of the factorization ($\mu_\mathrm{F}$) and renormalization ($\mu_\mathrm{R}$) scales~\cite{PhysRevLett.103.012002,JHEP:ttbb}, and are complicated by the presence of two very different scales in this process: the top quark mass and the jet transverse momentum (\pt).
Therefore, experimental measurements of production cross sections {$\Pp\Pp\to \ttjj\ (\sttjj)$ and $\Pp\Pp\to \ttbb\ (\sttbb)$ can
provide an important test of NLO quantum chromodynamics (QCD) theory calculations and important input for describing the main background in the search for the \ttH process.
Previous cross section and ratio measurements at $\sqrt{s}= 7$ and 8\TeV have been reported by the CMS~\cite{cms:ttbb8TeV,cms:TOP-12-041} and ATLAS Collaborations~\cite{atlas:ttbb8TeV}.

In this Letter, the measurements of the cross sections \sttbb and \sttjj and their ratio are presented using a data sample of pp collisions collected at a centre-of-mass energy of 13\TeV at the CERN LHC by the CMS experiment, and corresponding to an integrated luminosity of 2.3\fbinv~\cite{LUM-15-001}.
Events are selected with the final state consisting of two leptons ($\Pe$ or $\PGm$) and at least four reconstructed jets, of which at least two are identified as b quark jets.
The cross section ratio is measured with a smaller systematic uncertainty exploiting the partial cancellation of uncertainties.

\section{The CMS detector and event simulation}

The central feature of the CMS apparatus is a superconducting solenoid of 6\unit{m} internal diameter, providing a magnetic field of 3.8\unit{T}.
Within the solenoid volume are a silicon pixel and strip tracker, a lead tungstate crystal electromagnetic calorimeter (ECAL), and a brass and scintillator hadron calorimeter (HCAL), each composed of a barrel and two endcap sections. Forward calorimeters extend the pseudorapidity ($\eta$) coverage provided by the barrel and endcap detectors.
Muons are reconstructed in gas-ionization detectors embedded in the steel flux-return yoke outside the solenoid.
A more detailed description of the CMS detector, together with a definition of the coordinate system used and the relevant kinematic variables, can be found in Ref.~\cite{Chatrchyan:2008zzk}.

The Monte Carlo (MC) simulated samples for the \ttbar signal are generated by the \POWHEG~(v2) event generator~\cite{Nason:2004rx,Frixione:2007vw,Alioli:2010xd} at NLO, interfaced with \PYTHIA (v8.205)~\cite{pythia,pythia8}
using the tune CUETP8M1~\cite{pythiatune}
to provide the showering of the partons and
to match soft radiation with the contributions from the matrix elements (MEs).
The NNPDF3.0~\cite{Ball:2014uwa} set of the parton distribution functions (PDFs) is used.
The \MADGRAPH~(v5.1.5.11) event generator~\cite{MADGRAPHold2}
with MEs at leading order (LO), allowing up to three additional partons, including b quarks,
and the {\MGvATNLO}~(v2.2.2) event generator~\cite{amcnlo} are both used for cross-checks and studies of systematic uncertainties.
The \ttbar samples are normalized to the next-to-next-to-leading-order (NNLO) cross section calculation~\cite{Czakon:2011xx}.
The $\PW$+jets and $\cPZ/\gamma^{*}$+jets processes are simulated in {\MGvATNLO} and are normalized to their NNLO cross sections~\cite{Li:2012wna}.
The single top quark associated production with a $\PW$ boson ($\Pp\Pp \to \PQt\PW$ and $\Pp\Pp\to\PAQt\PW$) is simulated in the five-flavour scheme in \POWHEG(v1) at NLO and normalized to an approximate NNLO cross section calculation~\cite{kidonakis:tw},
while the $t$-channel single top quark events are simulated
in the four-flavour scheme in {\MGvATNLO}.
The multijet production is modelled in \PYTHIA with LO MEs.
The CMS detector response is simulated using {\GEANTfour}~(v9.4)~\cite{Agostinelli:2002hh}.
The events in simulation include the effects of additional interactions in the same or nearby bunch crossings (pileup) and are weighted according
to the vertex distribution observed in data.
The number of pileup interactions in data is estimated from the measured bunch-to-bunch instantaneous luminosity
and the total inelastic cross section~\cite{13TeVXsection}.

\section{Definition of signal events}

Measurements are reported for two different regions of the phase space: the visible and the full phase space.
The result in the visible phase space is measured at the particle level,
using the stable particles after the hadronization,
to reduce the possible theoretical and modelling uncertainties,
while the purpose of performing the result in the full phase space is
to facilitate comparisons to NLO calculations or measurements in other decay modes.

To define the visible phase space, all $\ttbb$ final-state particles
except the neutrinos,
\ie the charged leptons and jets originating from the decays of the top quarks,
as well as the two additional b quark jets (``b jets''), are required to be within the experimentally accessible kinematic region.
The leptons must have $\pt>20\GeV$, and $\abs{\eta}<2.4$.
Electrons or muons originating from the leptonic decays of $\Pgt$ leptons produced in $\PW \to \Pgt\Pgn$ decays are included.
The particle-level jets are obtained by combining all final-state particles, excluding neutrinos, at the generator level with an anti-\kt clustering algorithm~\cite{antikt}
with a distance parameter of 0.4 and are required to satisfy $\abs{\eta} < 2.5$ and $\pt > 20$\GeV,
which is lower than the reconstructed minimum jet \pt due to jet resolution
-- to have all events that pass the reconstructed jet \pt in the visible phase space.
Jets that are within $\DR =\sqrt{\smash[b]{(\Delta \phi)^2+(\Delta \eta)^2}} = 0.5$ units of an identified electron or muon are removed,
where $\Delta \phi$ and $\Delta \eta$ are the differences in azimuthal angle and pseudorapidity between
the directions of the jet and the lepton.
To identify the b and c quark jets (``c jets'') unambiguously, the b and c hadron momenta are scaled down to a negligible value
and included in the jet clustering (so called ``ghost matching")~\cite{Cacciari:2008gn}.
The b and c jets are then identified by the presence of the corresponding ``ghost'' hadrons
among the jet constituents.

Simulated events are categorized as coming from the \ttjj process
if they contain at least four particle-level jets, including
at least two jets originating from b quarks, and two leptons ($\ttjj \to \PQb\PWp\PAQb\PWm\mathrm{jj}\to \PQb{\ell}^{+} \Pgn\PAQb {\ell}^{-} \Pagn \mathrm{jj}$).
The \ttjj sample
contains four components
according to
the number of b and c jets in addition to the two b jets required from the
top quark decays. The four components are the \ttbb final state with two
b jets, the \ttb final state with one b jet and one lighter-flavour jet,
the \ttcc final state with two c jets, and the \ttll final state with two
light-flavour jets (from a gluon or u, d, or s quark) or one light-flavour
jet and one c jet.
The \ttb final state mainly originates from the merging of two b jets
or the loss of one of the b jets caused by the acceptance requirements.

\section{Event selection}
The events are recorded at $\sqrt{s} = 13$\TeV using a dilepton trigger~\cite{TRG-12-001} that requires the presence of two isolated leptons ($\Pe$ or $\PGm$)
both with \pt larger than 17\GeV.

The particle-flow (PF) event algorithm~\cite{PAS_PFT-09-001,PAS_PFT-10-001} reconstructs and identifies each individual particle with an optimized combination of information
from the various elements of the CMS detector.
The energy of photons is directly obtained from the ECAL measurement.
The energy of electrons is determined from a combination of the electron momentum at the primary interaction vertex
as determined by the tracker, the energy of the corresponding ECAL cluster, and
the energy sum of all bremsstrahlung photons spatially compatible with originating from the electron track.
The energy of muons is obtained from the curvature of the corresponding track
reconstructed by combining information from the silicon tracker and the muon system~\cite{paper:muon}.
The energy of charged hadrons is determined from a combination of their momenta measured in the tracker
and the matching ECAL and HCAL energy deposits, corrected for zero-suppression effects and for the response function of the calorimeters to hadronic showers.
Finally, the energy of neutral hadrons is obtained from the corresponding corrected ECAL and HCAL energy.

The leptons and all charged hadrons that are associated with jets are required to originate from the primary vertex, defined as the vertex with the highest $\sum{\pt^2}$ of its associated tracks.
Muon candidates are further required to have a high-quality fit including a minimum number of hits in both systems.
Requirements on electron identification variables based on shower shape and track-cluster matching are further applied
to the reconstructed electron candidates~\cite{Khachatryan:2015hwa, EGM-14-001,DP-2015_067}.
Muons and electrons must have $\pt > 20$\GeV and $\abs{\eta}< 2.4$.

To reduce the background contributions of muons or electrons from semileptonic heavy-flavour decays, relative isolation criteria are applied.
The relative isolation parameter, $I_\text{rel}$, is defined as the ratio of the summed \pt of all objects
in a cone of $\DR=0.3$ ($\DR=0.4$) units
around the electron (muon) direction to the lepton \pt.
Different cone sizes for electron and muon are used to maximize the sensitivity.
The objects considered are the
charged hadrons associated with the primary vertex as well as
the neutral hadrons and photons, whose energies are corrected to take into account pileup effects.
Thus,
\begin{equation}
I_\text{rel}=\frac{\sum{\pt^\text{charged hadron}} + \sum{\pt^\text{neutral hadron}} + \sum{\pt^\text{photon}}}{\pt^\text{lepton}}.
\end{equation}

The muon candidates are required to have $I_\text{rel}< 0.15$.
For the electron candidates, different $I_\text{rel}$ thresholds (0.077 or 0.068) are applied
depending on the pseudorapidity of the candidate ($\abs{\eta}< 1.48$ or $1.48 \leq \abs{\eta}< 2.40$).
These thresholds are obtained from a multivariate analysis technique and result from the considerable
differences in both the ECAL and the tracker in the two pseudorapidity regions.
The efficiencies for the above lepton identification requirements are
measured using $\Z$ boson candidates in data with a dilepton invariant mass between 70 and 130\GeV,
and are compared with the values from the simulation.
The differences between the two evaluations are applied as a correction to the simulation.

The event selection requires the presence of two isolated opposite-sign leptons of invariant mass $M_{\ell\ell} > 12$\GeV.
Lepton pairs of the same flavour (\EE, \MM) are rejected if their invariant mass is within 15\GeV of the $\Z$ boson mass.
The missing transverse momentum vector \ptvecmiss is defined as the projection on the plane perpendicular to
the beams of the negative vector sum of the momenta of all reconstructed PF candidates in the event.
Its magnitude is referred to as $\ptmiss$.
In the same-flavour channels, remaining backgrounds from $\Z$+jets processes are suppressed by demanding $\ptmiss> 30$\GeV.
For the \emu channel, no $\ptmiss$ requirement is applied.

Jets are reconstructed using the same anti-\kt clustering algorithm as particle-level jets in the simulations, with the PF candidates as input particles.
The jet momentum is determined as the vectorial sum of all PF candidate momenta in the jet and is found from simulation to be within 5 to 10\% of the true momentum over the whole \pt spectrum and detector acceptance. An offset correction is applied to jet energies to take into account the contribution from
pileup interactions.
Jet energy corrections are derived from simulation and confirmed with in situ measurements of the energy balance in dijet and photon+jet events~\cite{paper:jet}.
Additional selection criteria are applied to each event to remove spurious jet-like features originating from isolated noise patterns in certain HCAL regions.
The event must contain at least four reconstructed jets
with $\pt > 30$\GeV and $\abs{\eta}< 2.4$,
of which at least two jets must be identified as b jets, using the combined secondary vertex (CSV) algorithm (v2),
which combines secondary vertex information with lifetime information of single tracks to produce a b tagging discriminator~\cite{paper:btagid}.
A b tagging requirement on this discriminator is applied, which has an efficiency of about 60--70\% for b jets and
a misidentification probability of 1\% for light-flavour jets and 15--20\% for c-flavour jets~\cite{btagPASRun2}.

Differences in the b tagging efficiencies between data and simulation~\cite{paper:btagid} are accounted for by reweighting the shape of the CSV b tagging discriminator
distribution in the simulation to match that in the data.
Data/simulation \pt- and $\eta$-dependent correction factors are derived from the control samples
separately for light- and heavy-flavour jets, that are described in Section~\ref{sys}}.

The diboson, W+jets and multijet contributions are found to be negligible after the full event selection.
The $\Z$+jets background is estimated from data using control samples enriched in $\Z$ boson events.

Table~\ref{table:selfinal} gives the predicted number of events for each physics process
and for each lepton category,
as well as a comparison of the total number of
events expected from the simulation and observed in data.
Since the full event selection requires at least two b-tagged jets, a condition which is usually satisfied by \ttbar events, only 5\% of the events are from non-\ttbar processes.
The \ttb final state is predominantly composed of $\ttbb$ events
where there is one lost b jet due to acceptance requirements (73\% of \ttb events).
The background contribution from \ttbar events that fail the visible phase space requirements is labelled ``\ttbar others''.
The number of observed events with four or more reconstructed jets is lower than
the prediction from the simulation,
a condition that is also observed in the lepton+jets decay mode~\cite{cms:TOP-16-008}.

\begin{table*}[h!]
\topcaption{
Predicted number of events for each physics process and for each dilepton category, their total, and the observed number of events.
Results are shown after the final event selection.
The $\Z$+jets normalization and uncertainty are calculated from data, while all other predictions and statistical uncertainties come from the simulated data samples.
The \ttbar sample for event categorization is from the \POWHEG~(v2) event generator interfaced with \PYTHIA~(v8.205).
}
\centering
\newcolumntype{x}{D{,}{\,\pm\,}{3.2}}
\newcolumntype{y}{D{,}{\,\pm\,}{4.2}}

\begin{tabular}{l *{3}{x}y}
\hline
Process                          & \multicolumn{1}{l}{\EE} &   \multicolumn{1}{l}{\MM}  &  \multicolumn{1}{l}{\emu}        &  \multicolumn{1}{l}{All}  \\ \hline
\ttbb                            & 6.3 , 0.4         &   8.6 , 0.4      &  24  , 1       & 39 , 1  \\  \hline
\ttb                             & 16  , 1           &   21  ,1        &  57  , 2       & 95 , 2  \\
\ttcc                            & 7.7 , 0.4         &   11  , 1        &  27  , 1       & 46 , 1    \\
\ttll                            & 157 , 2           &   220 , 2        &  596 , 3       & 972 , 4  \\
\ttbar~ others                   & 18  , 1           &   19  , 1        &  61  , 1       & 99 , 1   \\
\ttbar V                   & 2.5 , 0.1         &   3.2 , 0.2      &  7.3 , 0.2     & 14 , 1  \\
Single \PQt                         & 6.6 , 0.8         &   8.4 , 0.8      &  23  , 2       & 39 , 2   \\
$\Z$+jets                        & \multicolumn{1}{l}{ $\hspace*{0.5ex}0.8^{+1.0}_{-0.8}$}         &   5.4 , 1.5      &  0.6 , 0.5     & 6.8 , 1.9   \\[0.2ex]
\hline
Total                            & 215 , 2           &   297 , 3        & 796 , 4        & 1311 , 6   \\
\hline
Data                             & \multicolumn{1}{l}{186} & \multicolumn{1}{l}{288} & \multicolumn{1}{l}{682}   & \multicolumn{1}{l}{1156}            \\
\hline
\end{tabular}
\label{table:selfinal}
\end{table*}

\section{Cross section measurements}\label{Measurement}
\begin{figure*}[bt]
\centering
    \includegraphics[width=0.45\textwidth]{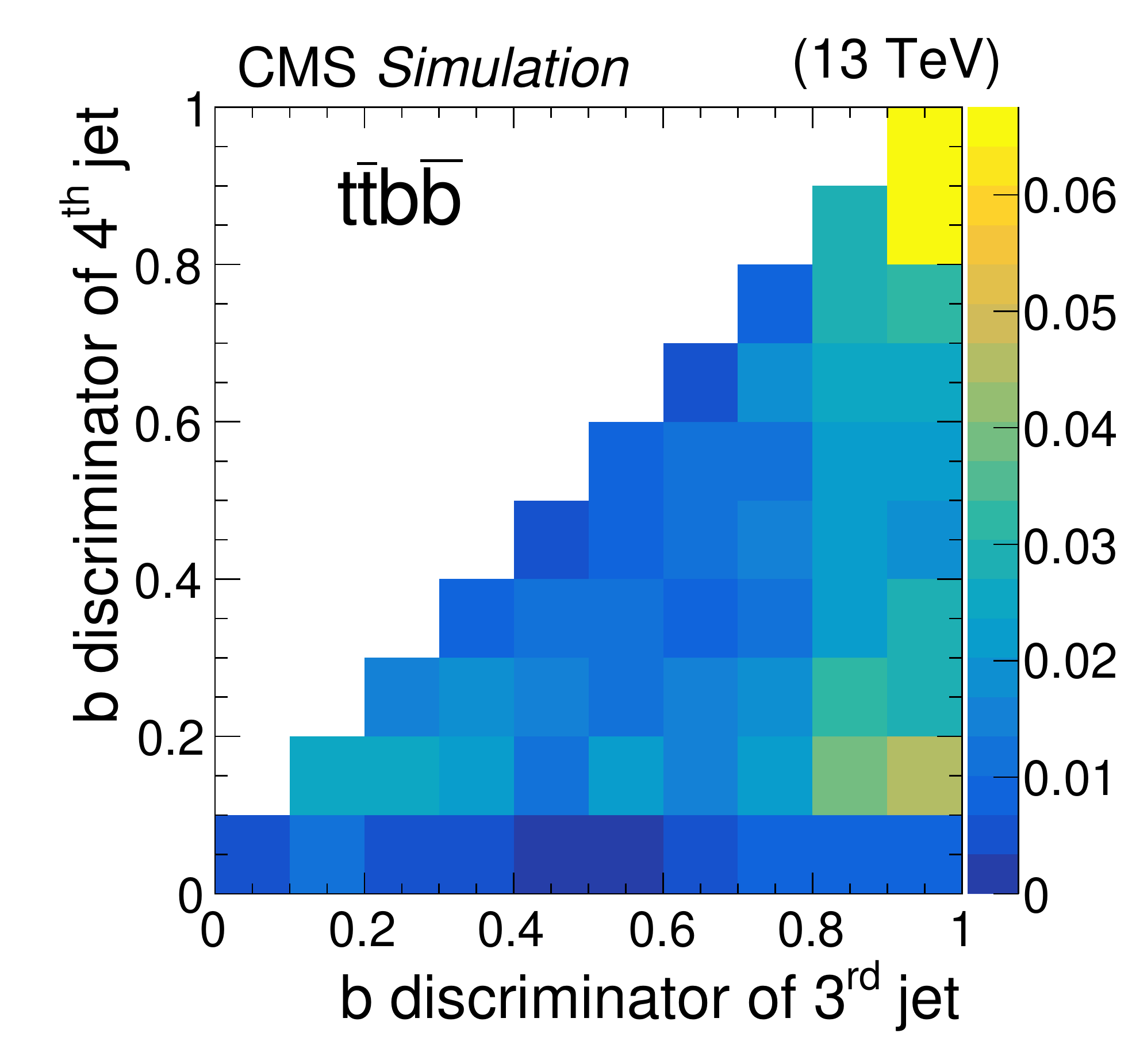}
    \includegraphics[width=0.45\textwidth]{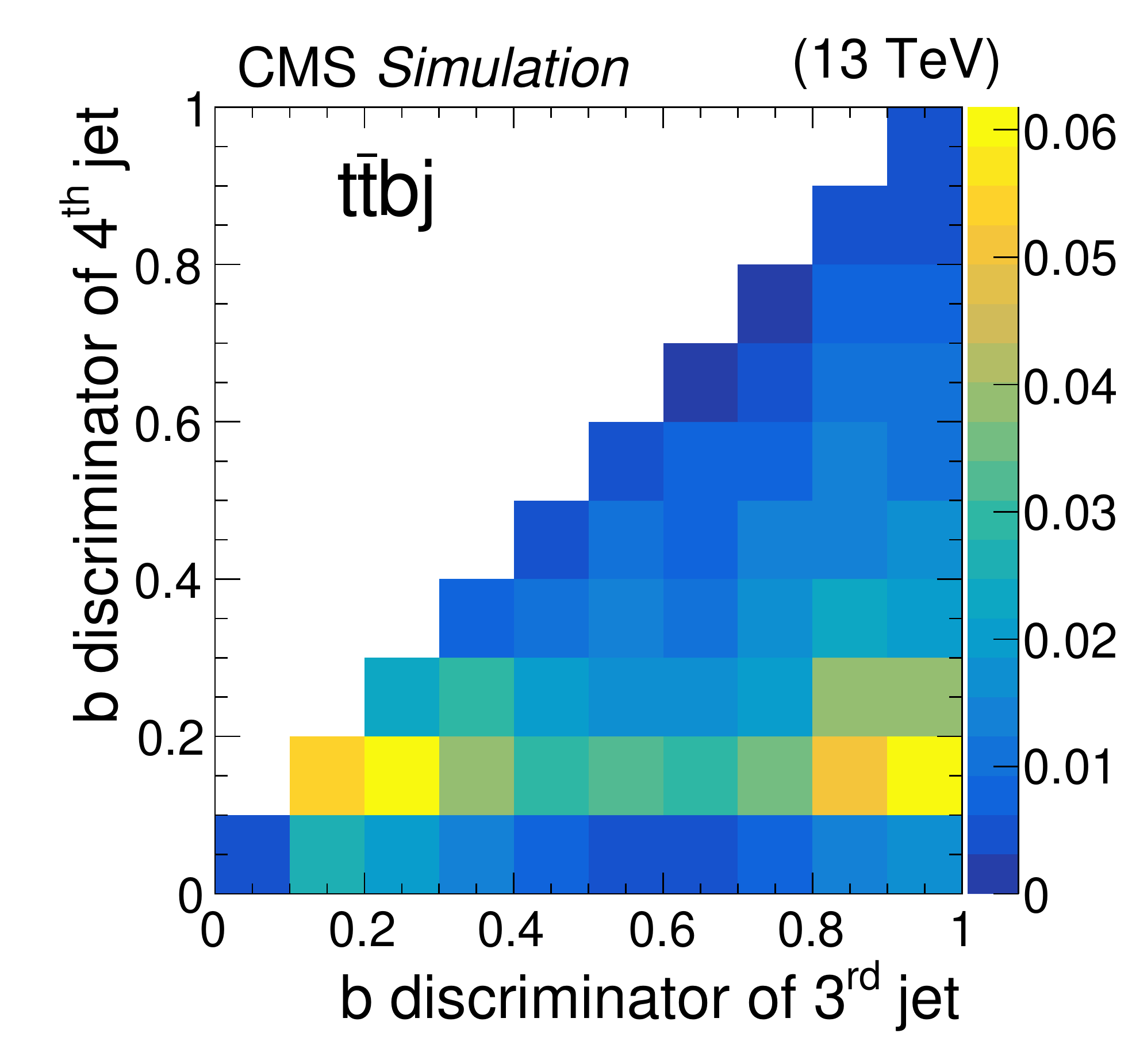}\\
    \includegraphics[width=0.45\textwidth]{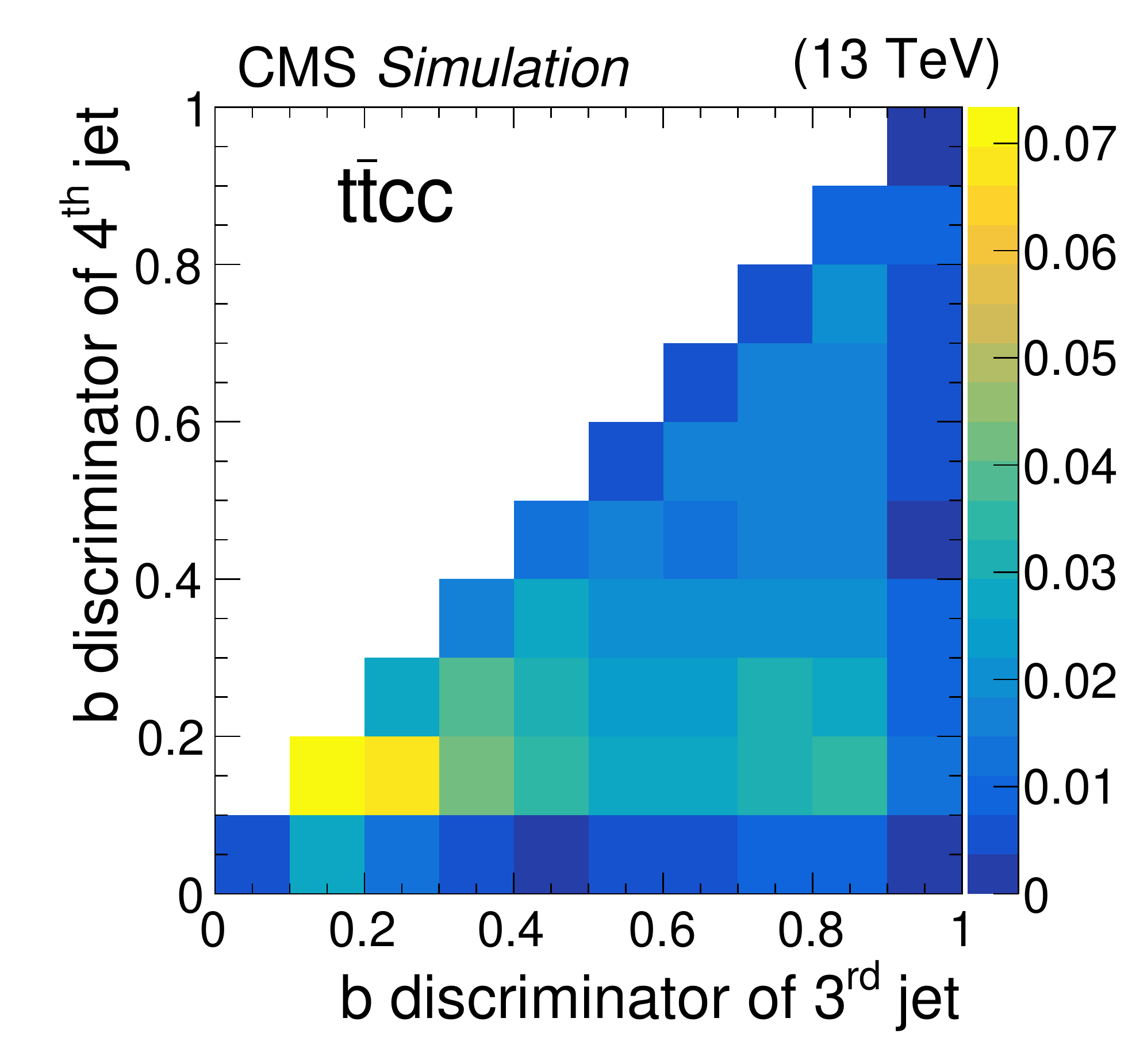}
    \includegraphics[width=0.45\textwidth]{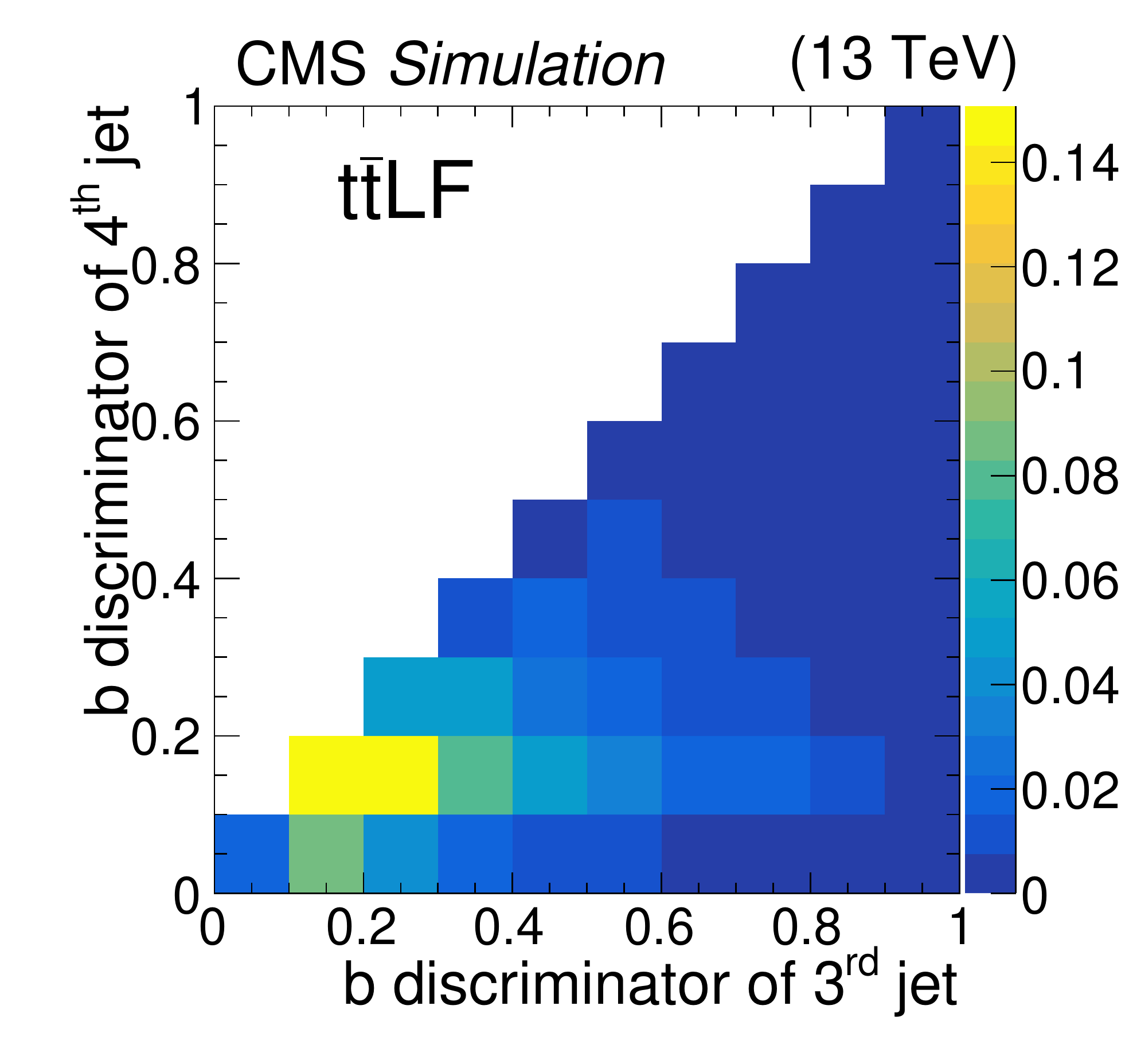}
    \caption{
Normalized 2D distributions of the b jet discriminators of the third
($x$-axis) and the fourth ($y$-axis) jets sorted in decreasing order of
b tagging discriminator value, after the full event selection
for $\ttbb$ (upper left), $\ttb$ (upper right), $\ttcc$ (lower left), and $\ttll$ (lower right) processes.
}
\label{fig:temp}
\end{figure*}

The first and the second jets in decreasing order of
the b tagging discriminator usually (in 85\% of \ttjj events) correspond to the b jets from the decays of top quarks,
and hence these jets provide no discriminating power between \ttbb and \ttjj events.
The third and the fourth jets
from \ttjj events are mostly light-flavour jets, while these are heavy-flavour jets for \ttbb events.
The normalized 2D distributions of the discriminators from simulation for the third and the fourth jets
are shown in Fig.~\ref{fig:temp}.
These 2D distributions are used to separate \ttbb events from other processes.
To extract the ratio of the number of \ttbb events to \ttjj events,
a binned maximum-likelihood fit is performed on the 2D distribution of the CSV b tagging discriminators
of the third and the fourth jets, where the three event categories $\ee$, $\emu$, and $\mumu$ are merged.

The number of \ttjj events and the ratio of the numbers of \ttbb events to \ttjj events are free parameters in the fit.
The $\ttcc$ and $\ttll$ processes have similar 2D distributions so their contributions are combined
based on the MC simulation.

The likelihood function is constructed as the product over all bins of a Poisson probability with a mean defined in each bin by
\ifthenelse{\boolean{cms@external}}{
\begin{multline}
\mathcal{M}(N_{\ttjj},R) = N_{\ttjj} \, \Bigl[ R \, \mathcal{F}_{\ttbb}^\text{norm}
\\+ R' \, \mathcal{F}_{\ttb}^\text{norm} + (1 - R - R' ) \, \mathcal{F}_{\ttll+\ttcc}^\text{norm} \Bigr]
\\+ N_{\ttjj} \, f_{\ttbar\text{ others}} + N_\text{bkg},
\end{multline}}{
\begin{equation}
\mathcal{M}(N_{\ttjj},R) = N_{\ttjj} \, \left[ R \, \mathcal{F}_{\ttbb}^\text{norm} + R' \, \mathcal{F}_{\ttb}^\text{norm} + (1 - R - R' ) \, \mathcal{F}_{\ttll+\ttcc}^\text{norm} \right] + N_{\ttjj} \, f_{\ttbar\text{ others}} + N_\text{bkg},
\end{equation}
}
where $\mathcal{F}_{\ttbb}^\text{norm}$, $\mathcal{F}_{\ttb}^\text{norm}$, and $\mathcal{F}_{\ttll+\ttcc}^\text{norm}$ are the normalized expectations for each bin of $\ttbb$, $\ttb$, and the combination of $\ttll$ and $\ttcc$, respectively.
The parameter $N_{\ttjj}$ denotes the number of the \ttjj events from the fit.
The quantity $f_{\ttbar\text{ others}}$ reflects the fraction of other $\ttbar$ processes in the \ttjj sample as calculated in simulation
($\ttbar\text{ others}$ divided by the sum of the \ttjj components in Table~\ref{table:selfinal}).
The other backgrounds, such as \ttbar V (V = W or Z) and single top quark processes are fixed to the simulation expectations, while the $\Z$+jets background is fixed to its estimation from control samples in data.
This remaining background not from the $\ttbar$ process is labelled $N_\text{bkg}$. The parameter $R$ is the ratio of the number of $\ttbb$ events with respect to the number of $\ttjj$ events, and $R'$ is the fraction of $\ttb$ events at the reconstruction level and constrained to the ratio of the number of the $\ttb$ events to $\ttbb$ events. It is fixed to 2.43 as calculated from the MC simulation (\POWHEG interfaced with \PYTHIA). The effect of this assumption is estimated as a systematic uncertainty in Section~\ref{sys}.
Values for $N_{\ttjj}$ of $950 \pm30$ events and $R$ of $0.056\pm0.008$ are obtained from the fit.
The correlation coefficient between the two parameters is 0.002.

The result obtained for $R$ is corrected to account for the different selection efficiencies for the two processes.
The event selection efficiencies, defined as the number of  \ttbb and \ttjj events after the full event selection
divided by the number of events
in the corresponding visible phase space, are 27\% and 12\%, respectively.
For the \ttbb process, there are at least 4 b jets in the events, therefore,
it is easier to fulfill the requirement of at least two b-tagged jets than the \ttjj process.

\begin{figure}
\centering
    \includegraphics[width=0.45\textwidth]{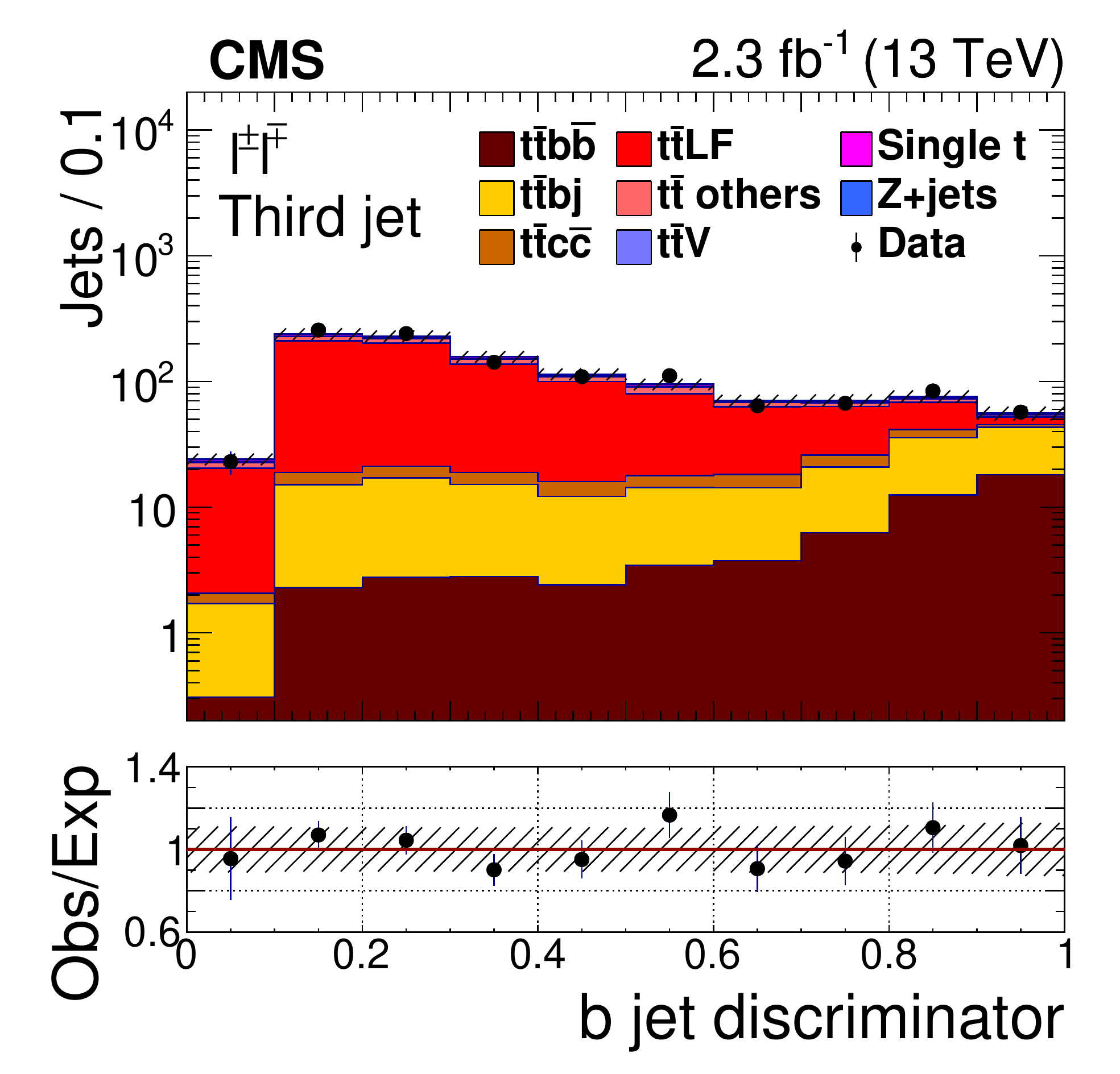}
    \includegraphics[width=0.45\textwidth]{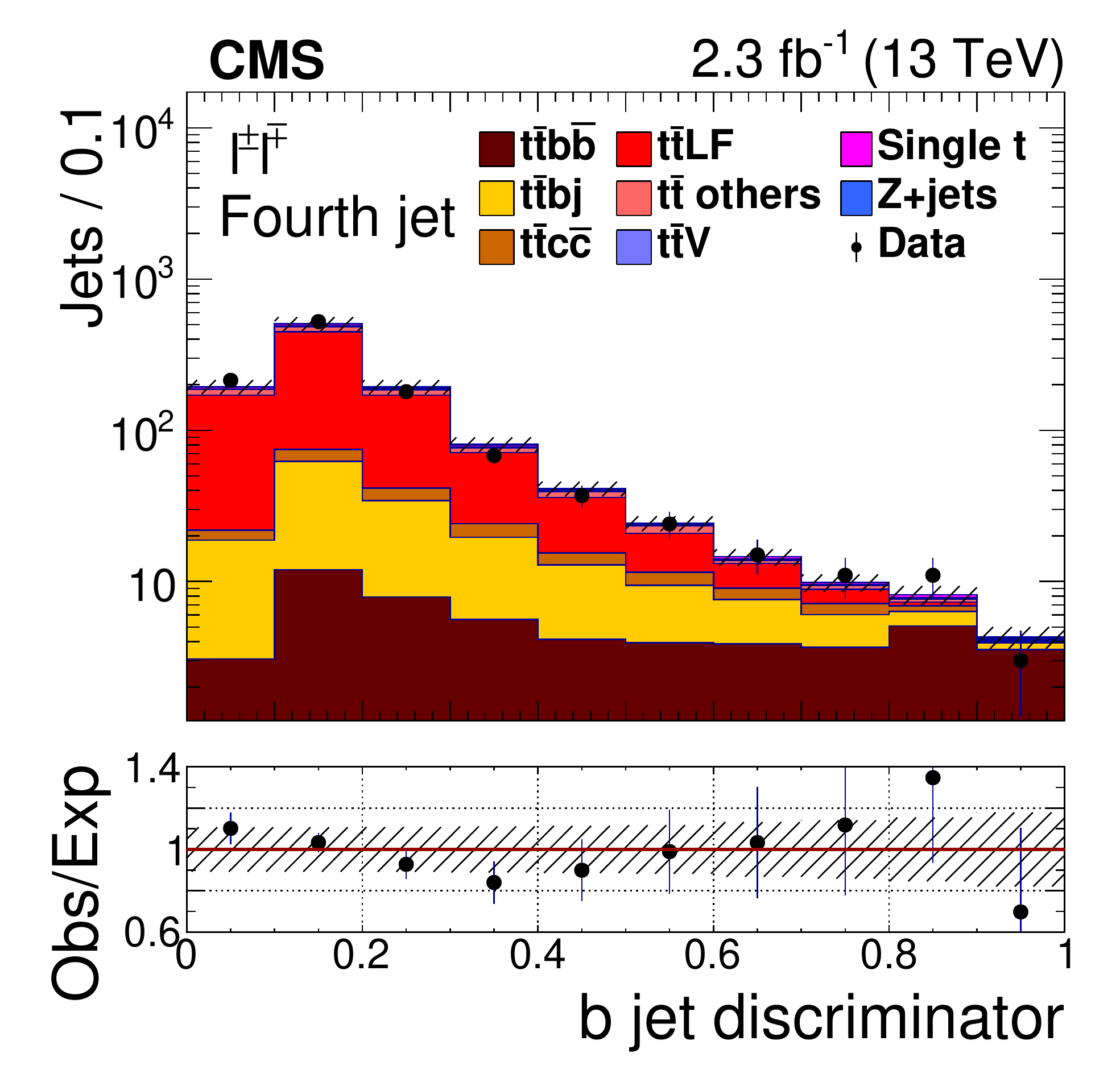}
    \caption{
Distributions of b jet discriminator for the third (\cmsLeft) and the fourth (\cmsRight) jets in decreasing order of b tagging discriminator value, after the full event selection.
The points show the data and the stacked histograms are from simulated events, normalized by the results of the fit.
The ratio of the number of data events to the expected number, as given by the stacked histograms, is shown in the lower panels.
The hatched region indicates the modelling uncertainty in the MC simulation.
}
\label{fig:fitresult}
\end{figure}

Figure~\ref{fig:fitresult} shows the comparisons of the b tagging discriminator distributions of the third
and the fourth jets in the events from data and simulation, where the
simulated histograms have been scaled to the fit result.

\begin{figure*}
\centering
     \includegraphics[width=0.45\textwidth]{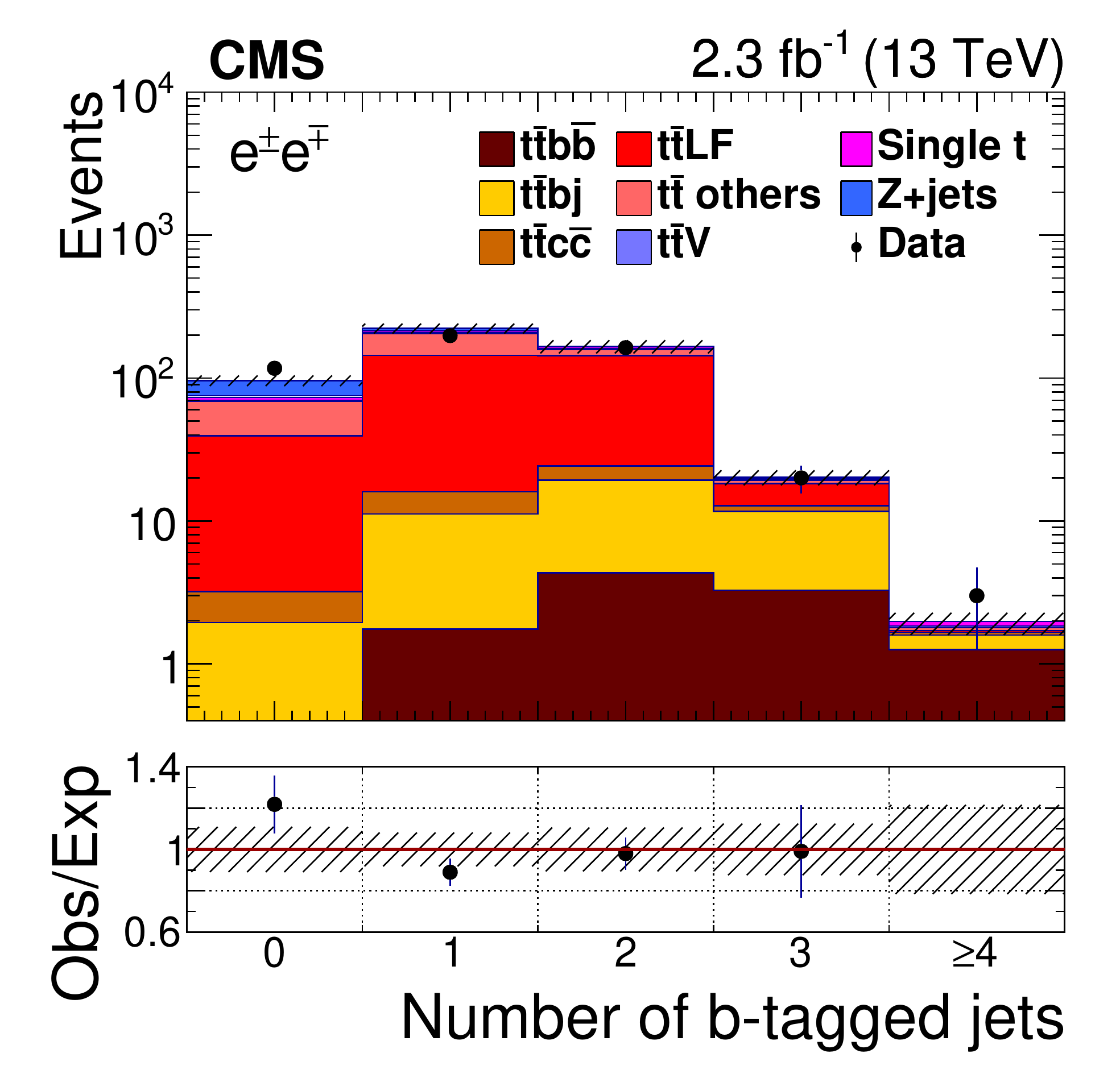}
     \includegraphics[width=0.45\textwidth]{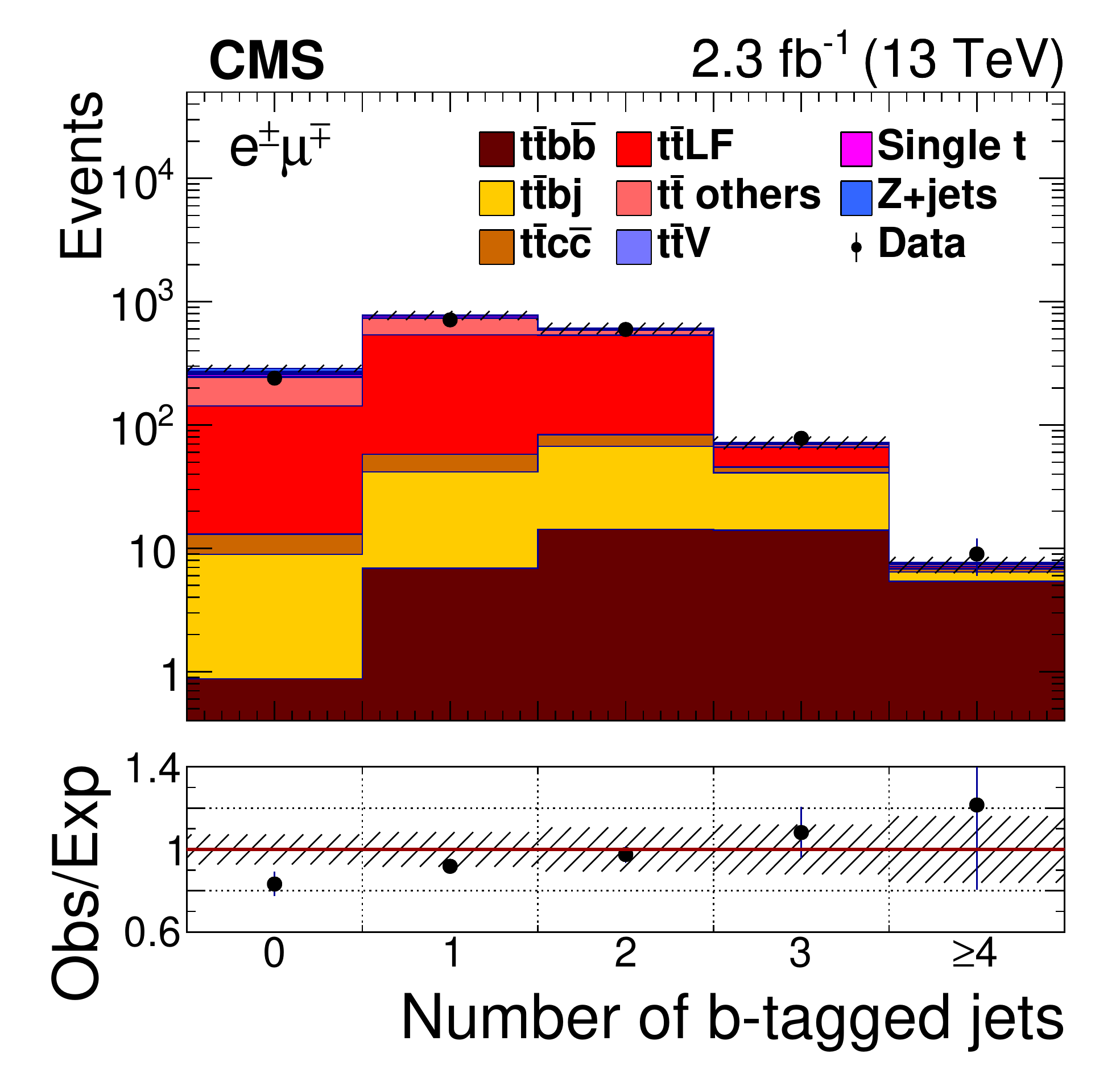}\\
     \includegraphics[width=0.45\textwidth]{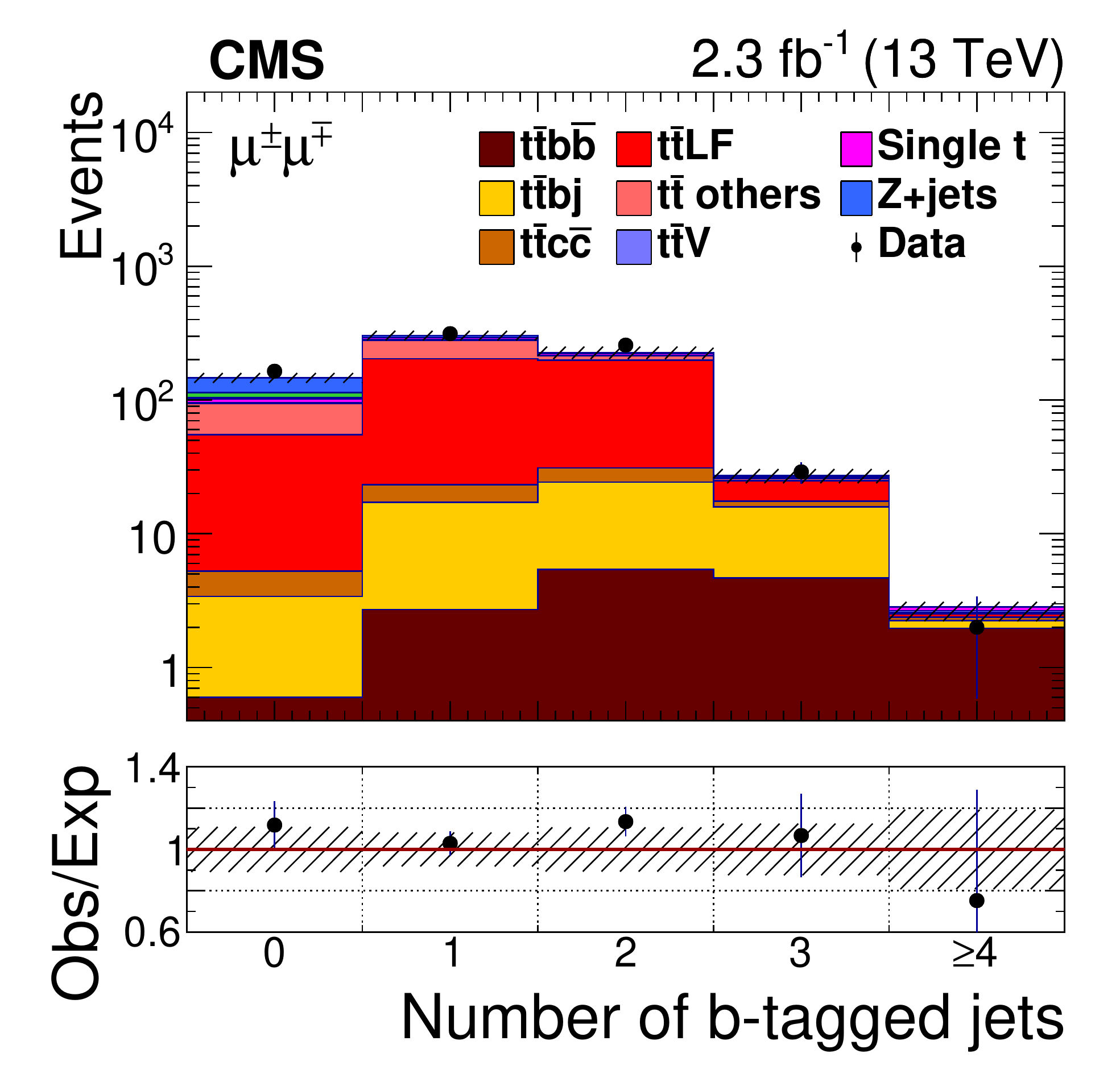}
     \includegraphics[width=0.45\textwidth]{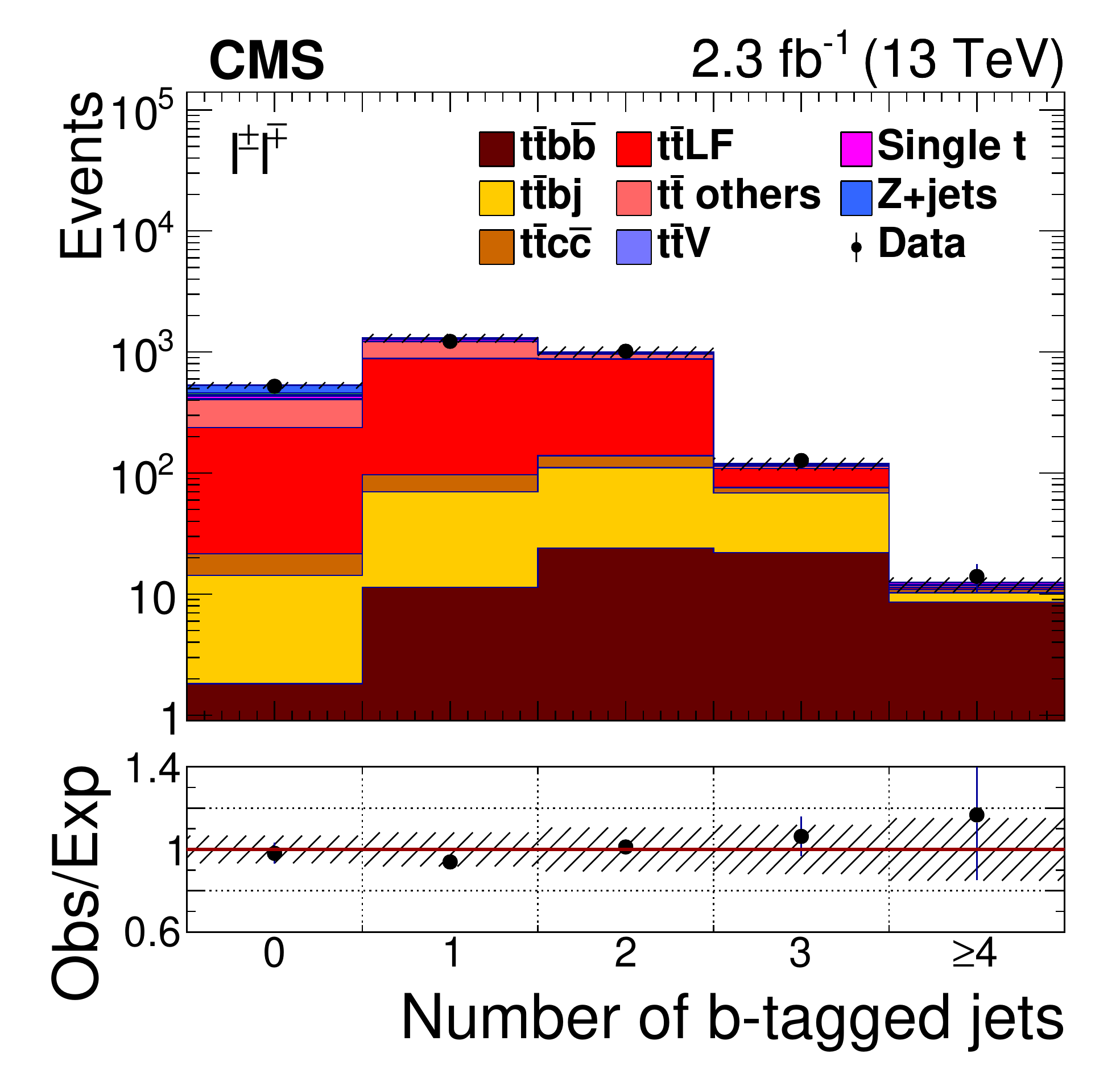}\\
    \caption{
Distribution of b jet multiplicity after the four-jet requirement, but without the b tagging requirement,
for the \ee (upper left), \emu (upper right), and \mumu (lower left) final states and the sum of the three final states (lower right).
The points show the data and the stacked histograms are from simulated events, normalized by the results of the fit.
The ratio of the number of data events to the expected number, as given by the stacked histograms, is shown in the lower panels.
The hatched region indicates the modelling uncertainty in the MC simulation.
}
    \label{fig:btag}
\end{figure*}

The b-tagged jet multiplicity distribution in Fig.~\ref{fig:btag} shows the comparison
between data and the simulation after the requirement of at least four jets, together
with the ratio of the number of data events to the expectation in the lower panel,
where the simulated histograms have been scaled to the fit result.

The \ttbb and \ttjj cross sections in the visible phase space are calculated using the relationship $\sigma_\text{visible} =  N / ( \epsilon  \mathcal{L} ) $,
where $\mathcal{L}$ is the integrated luminosity, $N$ is the number of events from the fit result,
and $\epsilon$ is the efficiency for each process.
For the purpose of comparing with the theoretical prediction and the measurements in the other decay modes, the cross sections in the full phase space are extrapolated from the cross sections in the visible phase space using the relation $\sigma_\text{full} = \sigma_\text{visible} / \mathcal{A}$, where $\mathcal{A}$ is the acceptance, defined as the number of events in the corresponding visible phase space divided by the number of events in the full phase space.
The acceptances are calculated based on the \POWHEG simulation and are 2.2\% and 2.0\% for \ttbb and \ttjj, respectively,
including the leptonic branching fraction of both W bosons~\cite{PDG2016}.
\section{Estimation of systematic uncertainties}\label{sys}
The systematic uncertainties are determined separately for the \ttbb and \ttjj cross sections, and their ratio.
In the ratio, many systematic effects cancel,
specifically normalization uncertainties, such as the ones related to the measurement of the integrated luminosity and
the lepton identification, including trigger efficiencies, since they are common to both processes.
The various systematic uncertainties in the measured values are
shown in Table~\ref{tab:sys} for the visible phase space.

\begin{table}[thb]
\centering
\topcaption{Summary of the systematic uncertainties in percentage (\%) from various sources contributing to $\sttbb$, $\sttjj$, and the ratio  $\sttbb$/$\sttjj$, for a jet \pt threshold of \pt $>$ 20 \GeV in the visible phase space.
 }
\label{tab:sys}
\begin{tabular}{@{}lccc@{}}
\hline
                        Source                          &   $\sttbb$  &  $\sttjj$   &  $\sttbb$/$\sttjj$   \\
                        \hline
                        b tag (b quark flavour)         &    19           &    4.7       &   19                    \\
                        b tag (c quark flavour)         &    14           &    1.3       &   14                   \\
                        b tag (light flavour)           &    14           &    9.8       &   9.7                   \\
                        JES \& JER                      &    7.8          &    7.4       &   2.6                   \\
                        Ratio of \ttbb and \ttb         &    2.6          &    0.5       &   2.6                   \\
                        Background modelling            &    3.8          &    3.5       &   1.6                   \\
                        \ttcc fraction in the fit       &    5.2          &    1.9       &   4.8                   \\
                        Lepton trigger/identification   &    3.0          &    3.0       &   0                   \\
                        Pileup                          &    0.4          &    $<$0.1   &   0.4                   \\
                        \hline
                        MC generator                    &    9.4           &    6.2      &   3.0                   \\
                        $\mu_\mathrm{F}$ and $\mu_\mathrm{R}$ scale  &    2.0           &    2.0      &   1.0                   \\
                        scale in PS                     &    13            &    9.9      &   10                   \\
                        PDFs                            &    0.5           &    0.5      &   $<$0.1                   \\
                        Efficiency (\ttcc fraction)     &    0            &    1.3      &   1.3                   \\
                        Jet multiplicity modelling       &    5.0          &    5.0       &   5.0                   \\
                        Simulation (statistical)                   &    1.5           &    1.5      &   1.5                   \\
                        Top quark \pt modelling                &    0.8           &    0.3      &   0.5                    \\
                        \hline
                        Integrated Luminosity                      &    2.3            &    2.3     &   0                   \\
                        \hline
                        Total uncertainty               &    34             &    19      &   28                    \\
\hline
\end{tabular}
\end{table}

The systematic uncertainties associated with the b tagging efficiency for heavy- and light-flavour jets are studied separately,
varying their values within the corresponding uncertainties.
The b-flavour correction factors are obtained using \ttbar enriched events by tagging one b jet and probing the other b jet.
Their dominant uncertainty comes from the contamination when one of the b jets is not reconstructed~\cite{cms:ttH} (indicated as ``b quark flavour'' in Table~\ref{tab:sys}).
The light-flavour jet correction factors are determined from $\Z$+jets enriched events with at least two jets (indicated as ``light flavour'' in Table~\ref{tab:sys}).
The uncertainty arises because in this control sample of $\Z$+jets, the contamination from the $\Z$+\bbbar process is not well modelled.
The correction factor for c jets is not measured, owing to the limited amount of data, and
is assumed to be unity with an uncertainty twice as large as for b jets~\cite{paper:btagid} (indicated as ``c quark flavour'' in Table~\ref{tab:sys}).
In the correction factor evaluation,
the statistical uncertainty, which can arise owing to low event yields in certain regions,
\eg at values of the b tagging discriminator near one, is also taken into account.

The b tagging discriminator can also be affected by the jet energy scale (JES) variations~\cite{paper:jet} since the efficiency correction changes through its \pt dependence.
The corresponding systematic uncertainty is obtained by varying the JES correction within its uncertainty
and repeating the whole analysis.
The uncertainty induced by the jet energy resolution (JER) is assessed by smearing the jet energy resolution in simulation
by an additional uncertainty dependent on $\eta$ of about 10\%~\cite{paper:jet}.

The ratio of $\ttbb$ events with respect to $\ttb$ events is based on the \POWHEG MC simulation.
The uncertainty arising from this rate is evaluated
by comparing the reference value (\POWHEG) with that of a {\MGvATNLO} sample, and \POWHEG samples
with different $\mu_\mathrm{F}$ and $\mu_\mathrm{R}$ scales in the ME and parton shower (PS) calculations.

The contributions from $\Z$+jets and single top quark processes are small,
and the 2D b tagging discriminator distributions
from these backgrounds are similar to those of the \ttll component.
Therefore, these backgrounds do not affect the measurement significantly.
The uncertainty caused by mismodelling of these backgrounds is assessed
by varying the contribution to cover the uncertainty in the single top quark production cross section (indicated as ``Background modelling'' in Table~\ref{tab:sys}).
An uncertainty to account for the modelling of the \ttcc fraction by simulations
is also assigned by varying the contribution by 50\% in the fit.
This is derived from the theoretical uncertainty on the \ttjj cross section.
For the efficiency of \ttjj events, the uncertainty owing to the heavy-flavour fraction is negligible because of their small fraction.

The systematic uncertainty in the lepton identification is
calculated by varying the correction factor for the efficiency within its uncertainty, as derived from
$\Z$ boson candidates as a function of lepton $\eta$ and $\pt$,
and also taking into account the different phase space between
$\Z$ boson and \ttbar events.

The systematic uncertainty in the number of pileup events is estimated by varying the total inelastic cross section by 5\% to cover all of the uncertainties
in the modelling of the pileup~\cite{13TeVXsection}.

The dependence of the correction factor at the particle level on the assumptions made in the MC simulation is another source of systematic uncertainty:
the generators \POWHEG and {\MGvATNLO} are compared and the difference in the efficiency is taken as systematic uncertainty.
The uncertainties from the $\mu_\mathrm{F}$ and $\mu_\mathrm{R}$ scales at the ME level are estimated
by making use of a weighting scheme implemented in \POWHEG to vary the scales by a factor of two up and down with respect to their reference values
$\mu_\mathrm{F}=\mu_\mathrm{R} = \sqrt{ \smash[b]{m_{\PQt}^2 + p_{\mathrm{T},\PQt}^2}}$, where $ m_{\PQt}$ is 172.5\GeV,
with $p_{\mathrm{T},\PQt}$ being the top quark transverse momentum.
The uncertainties from the $\mu_\mathrm{F}$ and $\mu_\mathrm{R}$ scales at the PS level
are assessed by using additional simulations where the scales are changed by a factor of two up and down relative to their reference values.
In simulation, event weights are calculated that represent the usage of the uncertainty eigenvector sets of the PDF.
The uncertainties in the PDFs are accounted for by using these various event weights.
The uncertainty from the modelling of jet multiplicity, in particular, the mismodelling for events with more than five jets, is also taken into account.
It is estimated to be 5\% by comparing the rates of high-multiplicity events in data and simulation.

The size of the MC sample used for \ttbb simulation being limited,
the uncertainty from the statistical fluctuations in the simulated event samples is assessed by repeating the fit with the method described in Ref.~\cite{MCstat}.
The difference of 1.5\% in the result is accounted for in the systematic uncertainty.

In addition to the theoretical and modelling uncertainties described above,
the uncertainty coming from the modelling of the top quark \pt distribution in the ME calculations is taken into account.
The uncertainty is calculated by taking the difference in shape between the parton-level \pt spectrum
from the ME generator and the unfolded \pt spectrum from the data~\cite{cms:topdiff8TeV}.
The uncertainty due to the top quark \pt modelling is negligible in this analysis, as shown in Table~\ref{tab:sys}.

Adding all these contributions in quadrature gives a total
systematic uncertainty of 28\% in the cross section ratio,
with the dominant contributions coming from the b tagging efficiency and the misidentification of light- and c-flavoured partons, followed by the matching scale systematic uncertainties.

The uncertainty in \sttjj is significantly smaller than that in \sttbb since the measurement of the latter requires the identification of multiple b jets.
The uncertainty in \sttbb is larger than that for the cross section ratio, since the uncertainties that are common between \ttbb and \ttjj, such as the JES uncertainty, partially or completely cancel in the ratio.

When extrapolating the measurements from the visible phase space to the full phase space,
the systematic uncertainty in the acceptance is included.
The effect of the MC modelling of the acceptance is estimated by comparing the results between {\MGvATNLO} and \POWHEG.
This uncertainty amounts to 4\% for each of the cross section measurements and 1\% for the cross section ratio.

\section{Results}
After accounting for all corrections and systematic effects,
the cross section ratio \sttbb/\sttjj\ is measured in the visible phase space
from a fit to the measured CSV b tagging discriminator distributions.
The measured cross section ratio in the visible phase space for events with particle-level jets is
\begin{equation}\label{eq:20GeVvisible}
    {(\sttbb/\sttjj)}^\text{vis}  = 0.024 \pm 0.003\stat \pm 0.007\syst.
\end{equation}
The result is obtained in the visible phase space, defined as events having two leptons with
$\pt > 20\GeV$ and $\abs{\eta}<2.4$, plus at least four jets, including at least two b jets with $\pt>20\GeV$ and $\abs{\eta}<2.5$.
The cross section ratio in the full phase space that uses the acceptance correction described in Section~\ref{Measurement} is
\begin{equation}\label{eq:20GeVfull}
    \sttbb/\sttjj = 0.022 \pm 0.003\stat \pm 0.006\syst.
\end{equation}
The predicted values from \POWHEG are $0.014 \pm0.001$ and $0.012 \pm0.001$ for the visible and full phase space, respectively, where
the uncertainty in the simulation is the sum in quadrature of
the statistical, and the $\mu_\mathrm{F}$/$\mu_\mathrm{R}$ scale systematic uncertainties.
The prediction obtained from \POWHEG simulation (interfaced with \PYTHIA) underpredicts the measured cross section ratio by a factor of 1.8, but it is compatible with the observation within two standard deviations.
The measured cross sections in the visible and the full phase space are presented in Table~\ref{tab:ttbbttjjX}.

\begin{table*}[htb]
\centering
\topcaption{
The measured cross sections \sttbb\ and \sttjj\ and their ratio
for the visible and the full phase space, corrected for acceptance and
branching fractions.
The uncertainties on the measurements show separately the statistical and systematic components,
while those are combined for the \POWHEG predictions.
}
\label{tab:ttbbttjjX}
\begin{tabular}{l|l|ccc}
\hline
\multicolumn{2}{c}{Phase space} & $\sttbb$ [pb] & $\sttjj$ [pb] & \sttbb/\sttjj \\ \hline
\multirow{2}{*}{Visible}  &   Measurement  & 0.088 $\pm$ 0.012 $\pm$ 0.029 & 3.7 $\pm$ 0.1 $\pm$ 0.7  & 0.024 $\pm$ 0.003 $\pm$ 0.007   \\
& SM (\POWHEG)  & 0.070 $\pm$ 0.009 & 5.1 $\pm$ 0.5 & 0.014 $\pm$ 0.001  \\ \hline
\multirow{2}{*}{Full}  &    Measurement    & 4.0   $\pm$ 0.6   $\pm$ 1.3   & 184  $\pm$ 6   $\pm$ 33  & 0.022 $\pm$ 0.003 $\pm$ 0.006  \\
& SM (\POWHEG)  & 3.2   $\pm$ 0.4   & 257 $\pm$ 26  & 0.012 $\pm$ 0.001  \\  \hline
\end{tabular}
\end{table*}

\section{Summary}
Measurements of the cross sections \sttbb and \sttjj and their ratio \sttbb/\sttjj are presented using a data sample recorded in pp collisions at $\sqrt{s} = 13$\TeV, corresponding to an integrated luminosity of 2.3\fbinv.
The cross section ratio has been measured in a visible phase space region using the dilepton decay mode of \ttbar events and corrected to the particle level, corresponding to the detector acceptance.
The measured cross section ratios in the visible and the full phase space are $\sttbb/\sttjj = 0.024 \pm 0.003\stat \pm 0.007\syst$ and $\sttbb/\sttjj = 0.022 \pm 0.003\stat \pm 0.006\syst$, respectively, where a minimum transverse momentum for the particle-level jets of 20\GeV is required.
The \ttH contribution, being negligible, is not removed from data.
Theoretical ratios predicted from the \POWHEG simulation (interfaced with \PYTHIA) are $0.014\pm0.001$ for the visible and $0.012\pm0.001$ for the full phase space, which are lower than the measured values but consistent within two standard deviations.
The individual cross sections
$\sttbb = 4.0 \pm 0.6\stat\pm1.3\syst$\unit{pb} and $\sttjj = 184 \pm 6\stat \pm 33\syst$\unit{pb}
have also been measured.
These results, in particular the ratio of the cross sections,
provide important information for the \ttH search,
permitting the reduction of a dominant systematic uncertainty
that derives from the uncertainty in the \ttbb background.
They can also be used as a figure of merit for
testing the validity of
next-to-leading-order QCD calculations at $\sqrt{s} = 13$\TeV.

\begin{acknowledgments}
We congratulate our colleagues in the CERN accelerator departments for the excellent performance of the LHC and thank the technical and administrative staffs at CERN and at other CMS institutes for their contributions to the success of the CMS effort. In addition, we gratefully acknowledge the computing centers and personnel of the Worldwide LHC Computing Grid for delivering so effectively the computing infrastructure essential to our analyses. Finally, we acknowledge the enduring support for the construction and operation of the LHC and the CMS detector provided by the following funding agencies: BMWFW and FWF (Austria); FNRS and FWO (Belgium); CNPq, CAPES, FAPERJ, and FAPESP (Brazil); MES (Bulgaria); CERN; CAS, MoST, and NSFC (China); COLCIENCIAS (Colombia); MSES and CSF (Croatia); RPF (Cyprus); SENESCYT (Ecuador); MoER, ERC IUT, and ERDF (Estonia); Academy of Finland, MEC, and HIP (Finland); CEA and CNRS/IN2P3 (France); BMBF, DFG, and HGF (Germany); GSRT (Greece); OTKA and NIH (Hungary); DAE and DST (India); IPM (Iran); SFI (Ireland); INFN (Italy); MSIP and NRF (Republic of Korea); LAS (Lithuania); MOE and UM (Malaysia); BUAP, CINVESTAV, CONACYT, LNS, SEP, and UASLP-FAI (Mexico); MBIE (New Zealand); PAEC (Pakistan); MSHE and NSC (Poland); FCT (Portugal); JINR (Dubna); MON, RosAtom, RAS, RFBR and RAEP (Russia); MESTD (Serbia); SEIDI, CPAN, PCTI and FEDER (Spain); Swiss Funding Agencies (Switzerland); MST (Taipei); ThEPCenter, IPST, STAR, and NSTDA (Thailand); TUBITAK and TAEK (Turkey); NASU and SFFR (Ukraine); STFC (United Kingdom); DOE and NSF (USA).

\hyphenation{Rachada-pisek} Individuals have received support from the Marie-Curie program and the European Research Council and Horizon 2020 Grant, contract No. 675440 (European Union); the Leventis Foundation; the A. P. Sloan Foundation; the Alexander von Humboldt Foundation; the Belgian Federal Science Policy Office; the Fonds pour la Formation \`a la Recherche dans l'Industrie et dans l'Agriculture (FRIA-Belgium); the Agentschap voor Innovatie door Wetenschap en Technologie (IWT-Belgium); the Ministry of Education, Youth and Sports (MEYS) of the Czech Republic; the Council of Science and Industrial Research, India; the HOMING PLUS program of the Foundation for Polish Science, cofinanced from European Union, Regional Development Fund, the Mobility Plus program of the Ministry of Science and Higher Education, the National Science Center (Poland), contracts Harmonia 2014/14/M/ST2/00428, Opus 2014/13/B/ST2/02543, 2014/15/B/ST2/03998, and 2015/19/B/ST2/02861, Sonata-bis 2012/07/E/ST2/01406; the National Priorities Research Program by Qatar National Research Fund; the Programa Clar\'in-COFUND del Principado de Asturias; the Thalis and Aristeia programs cofinanced by EU-ESF and the Greek NSRF; the Rachadapisek Sompot Fund for Postdoctoral Fellowship, Chulalongkorn University and the Chulalongkorn Academic into Its 2nd Century Project Advancement Project (Thailand); and the Welch Foundation, contract C-1845. \end{acknowledgments}
\bibliography{auto_generated}

\cleardoublepage \appendix\section{The CMS Collaboration \label{app:collab}}\begin{sloppypar}\hyphenpenalty=5000\widowpenalty=500\clubpenalty=5000\textbf{Yerevan Physics Institute,  Yerevan,  Armenia}\\*[0pt]
A.M.~Sirunyan, A.~Tumasyan
\vskip\cmsinstskip
\textbf{Institut f\"{u}r Hochenergiephysik,  Wien,  Austria}\\*[0pt]
W.~Adam, E.~Asilar, T.~Bergauer, J.~Brandstetter, E.~Brondolin, M.~Dragicevic, J.~Er\"{o}, M.~Flechl, M.~Friedl, R.~Fr\"{u}hwirth\cmsAuthorMark{1}, V.M.~Ghete, N.~H\"{o}rmann, J.~Hrubec, M.~Jeitler\cmsAuthorMark{1}, A.~K\"{o}nig, I.~Kr\"{a}tschmer, D.~Liko, T.~Matsushita, I.~Mikulec, D.~Rabady, N.~Rad, H.~Rohringer, J.~Schieck\cmsAuthorMark{1}, J.~Strauss, W.~Waltenberger, C.-E.~Wulz\cmsAuthorMark{1}
% ...
\vskip\cmsinstskip
1:~~Also at Vienna University of Technology, Vienna, Austria\\
% ...

\end{sloppypar}
\end{document}